\documentclass[twocolumn,prl,amsmath,amssymb]{revtex4}

\usepackage{graphicx}
\usepackage{dcolumn}
\usepackage{bm}
\newcolumntype{.}{D{.}{.}{-1}}

\begin{document}

\preprint{APS/123-QED}

\title{Coherent optical phase transfer over a 32-km fiber with 1-s instability at $10^{-17}$}

\author{Seth M. Foreman$^1$, Andrew D. Ludlow$^1$, Marcio~H.~G.~de~Miranda$^1$, Jason E. Stalnaker$^2$,
Scott A. Diddams$^2$, and Jun Ye$^1$} \affiliation{$^1$JILA,
National Institute of Standards and Technology and University of
Colorado \\Department of Physics, University of Colorado, Boulder,
CO 80309-0440} \affiliation{$^2$Time and Frequency Division, MS
847, National Institute of Standards and Technology, Boulder,
Colorado 80305, USA}

\date{\today}

\begin{abstract}
The phase coherence of an ultrastable optical frequency reference
is fully maintained over actively stabilized fiber networks of
lengths exceeding 30 km. For a 7-km link installed in an urban
environment, the transfer instability is $6 \times 10^{-18}$ at
1-s. The excess phase noise of 0.15 rad, integrated from 8 mHz to
25 MHz, yields a total timing jitter of 0.085 fs. A 32-km link
achieves similar performance. Using frequency combs at each end of
the coherent-transfer fiber link, a heterodyne beat between two
independent ultrastable lasers, separated by 3.5 km and 163 THz,
achieves a 1-Hz linewidth.

\end{abstract}


\maketitle

Optical atomic clocks with superior stability and accuracy
\cite{Bergquist2,BoydPRL} demand frequency transfer networks of
unprecedented stability for signal distribution, remote
synchronization and intercomparison. Clock signal transfer via
optical fiber networks has emerged as a promising solution
\cite{ForemanRSI} when phase noise in the long transmission path has
been effectively cancelled
\cite{Vessot,BergquistFermi,MaFiberCancellation}. In the microwave
domain, signals in the form of amplitude modulation of an optical
carrier have been transmitted over an 86-km optical fiber network,
where active stabilization of the fiber's group delay allows a
transfer instability of $5 \times 10^{-15}$ at 1-s and $2 \times
10^{-18}$ after 1 day \cite{Narbonneau}. However, a direct transfer
of the optical carrier itself \cite{ye03,Grosche} is destined to
achieve better stability, relying on the same advantage of high
spectral resolution as the optical clocks. Long-distance coherent
transfer of an ultrastable optical carrier signal, along with
frequency-comb-based optical coherence distribution across the
entire visible spectrum \cite{Ludlow1,Coddington}, permits a wide
variety of applications. They include phase-coherent large arrays of
radio telescopes \cite{Musha}, precisely synchronized advanced light
sources based on large-scale accelerators \cite{Schoenlein,
Cavalieri}, and precision optical interferometry over a long
distance or encircling a large area.

In this Letter, we report experimental implementations of a fully
coherent ($<$1 rad of accumulated optical phase noise)
optical-frequency-distribution fiber system spanning tens of
kilometers in distance and hundreds of nanometers in optical
spectrum.  While 1-s instability of $6 \times 10^{-17}$ has been
achieved in shorter ($< 1$-km) links \cite{Coddington}, the $2
\times 10^{-17}$ instability achieved here on a $> 10$-km link is
$\sim2$ orders of magnitude lower than previously published results
\cite{Musha, Grosche}. In a 7-km fiber network installed in an urban
environment, the transmission instability is reduced to $6 \times
10^{-18}$ at 1 s and $1 \times 10^{-19}$ at $10^5$ s, limited by the
out-of-loop measurement scheme. The accumulated phase noise,
integrated from 8 mHz to 25 MHz, is 0.15 rad, representing a
coherent optical transfer for timescales substantially longer than
the coherence times of the current best optical references. An
extended 32-km link employs a transceiver configuration, which is an
important first step towards the realization of coherent repeaters
for unlimited distribution distances. The system achieves similar
stability performance. Using frequency combs at each end of a third
fiber link, we remotely compare two Hz-linewidth lasers separated by
3.5 km of fiber and spectrally separated by 163 THz.  The effective
heterodyne optical beat has a 1-Hz linewidth, limited by the
ultrastable lasers.

\begin{figure}[t]
\resizebox{8.5cm}{!}{
\includegraphics[angle=0]{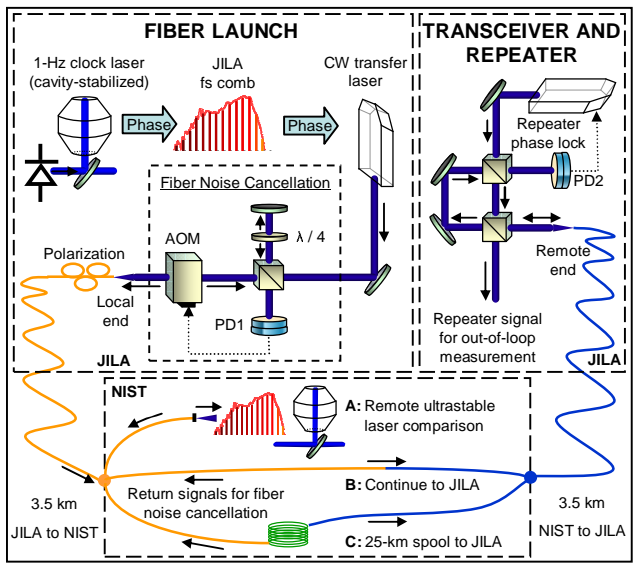}}
\caption{\label{Fig1}(color online) Schematic for phase-coherent
transfer of an optical carrier.  At the Fiber Launch the cw transfer
laser can be stabilized to a sub-Hz-linewidth clock laser via a fs
frequency comb, or left free-running. The Fiber Noise Cancellation
uses a double-pass AOM as the actuator.  Round-trip light used
in-loop for the noise cancellation comes from either a
transceiver/repeater laser (transceiver method), or from a partially
reflective gold coating on the flat tip of the fiber (reflection
method), at the remote
end of the network.  Fiber networks of three different lengths are characterized.  
}
\end{figure}

Figure~\ref{Fig1} shows the experimental scheme for phase-coherent
optical carrier transfer. To eventually compare the $^{87}$Sr
optical lattice clock \cite{BoydPRL} at JILA to other optical
clocks located at NIST \cite{Bergquist2, Oates}, we transmit the
light from a cw 1064-nm laser, which can be directly measured with
Ti:sapphire-based frequency combs serving as the optical clockwork
at both locations. For active cancellation of the fiber network's
phase noise, the transfer laser's coherence time must be longer
than $T_{rt}$, the fiber network's round-trip time. The transfer
laser used here has an intrinsic linewidth of 1 kHz in 1 ms,
sufficient for fiber noise cancellation over the 7-km link. For
longer fiber links and the remote ultrastable laser comparison,
the transfer laser must be stabilized to a sub-Hz linewidth 698-nm
laser\cite{Ludlow2} serving as the $^{87}$Sr clock's local
oscillator.  A self-referenced, octave-spanning Ti:sapphire laser
is used to transfer the clock laser's phase-stability across the
148-THz spectral gap to the transfer laser.

For the fiber noise cancellation, $\sim$1-mW from the transfer laser
is picked off by a polarizing beamsplitter (PBS) and detected on
photodiode PD1, while $\sim$40 mW is coupled into the fiber (below
threshold for stimulated Brillouin scattering) after being
frequency-shifted by an acousto-optic-modulator (AOM). Light
returning from the fiber network's remote end accumulates a
round-trip phase, again passes through the AOM, and is heterodyned
with the local light on PD1. The beat frequency $f_{\mathrm{fnc}}$
is used for fiber noise cancellation by phase-locking it to a
radio-frequency (rf) reference via active feedback to the AOM's
driving frequency \cite{BergquistFermi,MaFiberCancellation}. Noise
processes that are stationary during $T_{rt}$ (noise bandwidth $< 1
/ 2\pi T_{rt}$) are therefore pre-compensated by the AOM, whereas
noise at frequencies above this bandwidth is uncancelled. Due to
burst noise in the fiber link, digital pre-scaling of
$f_{\mathrm{fnc}}$ by a division of 50 is used to give the phase
lock enough dynamic range to avoid cycle slips.

We test two methods for returning light from the fiber's remote
end for noise cancellation.  The simpler technique (reflection
method) relies on partially reflective gold coating applied to the
flat-polished remote fiber tip.  The gold film transmits
$\sim10\%$ of the incident 1064-nm light, and reflects the rest
back to the launch port. The transmitted light is used for
out-of-loop measurements of the stabilized fiber network.  A more
complex technique (transceiver method) uses an angle-polished
remote fiber tip and an independent cw laser (transceiver laser)
operating at the remote end.  Light transmitted one-way through
the fiber network is heterodyned with the transceiver laser on
PD2, and the resulting beat signal is used to phase lock the
transceiver laser to the transmitted light with a controllable
constant frequency offset. Then $\sim$40 mW of light from the
transceiver laser is launched into the fiber network's remote end,
to be used for the in-loop fiber noise cancellation at the
launching end. Additional power from the transceiver laser is used
as a repeater signal for out-of-loop measurements of the
stabilized fiber network. The reflection method is simple but can
be corrupted by unwanted reflections at interconnections along the
fiber network. The more-complex transceiver method avoids this
problem since a different frequency is returned than the input
frequency.  Also, the transceiver method boosts the power at the
remote end and already realizes the setup necessary for a repeater
station in a much longer fiber network.

Polarization considerations for both methods are essential. A set
of fiber polarization paddles is used on the input end of the
fiber link to adjust the polarization of the round-trip light to
maximize the power of $f_{\mathrm{fnc}}$. The power still
fluctuates by $\sim1$~dB at frequencies of a few Hz, and slowly
degrades by as much as 5 to 10~dB over the course of one day due
to time-changing birefringence of the fiber network. Every few
hours the polarization paddles are adjusted to maximize the
strength of $f_{\mathrm{fnc}}$.  Another concern is fast
time-changing polarization mode dispersion \cite{Rashleigh}. For
the transceiver method it is necessary to ensure that the
polarization used to stabilize the transceiver laser is the same
as that it returns through the fiber network, so linear polarizers
are placed immediately outside both ends of the fiber. Without the
polarizers the fiber noise cancellation can not even be locked,
but with them it can be maintained indefinitely with even less
sensitivity to the fiber's changing birefringence than the
reflection scheme. Rotating the polarization paddles to degrade
the power in $f_{\mathrm{fnc}}$ by as much as 10~dB does not
affect the quality of the lock.

The bottom panel of Fig.~\ref{Fig1} shows how the fiber network can
be configured in three lengths for various measurements. Two 3.5-km
fibers in the Boulder Research and Administrative Network (BRAN)
connect JILA with NIST \cite{ye03}.  In setup A, only one
noise-cancelled fiber (reflection method) is used to transmit light
for comparing the clock laser at JILA to an independent ultrastable
laser at NIST \cite{YoungNarrowLaser} serving as the Hg$^+$ clock
laser.  A second octave-spanning frequency comb at NIST
\cite{FortierOctaveNIST} spans the 15-THz spectral gap between the
transfer laser and the 1126-nm clock laser. A heterodyne beat
between the transfer laser and one mode of the frequency comb at
NIST is used as the effective beat between the two remotely-located
clock lasers. In setup B, the two 3.5-km fibers are connected
together to form a single 7-km (one-way) fiber network. The 7-km
fiber has the local and remote ends located on the same table,
allowing direct out-of-loop measurement of the transfer system when
operated with either the reflection method or the transceiver
method. In setup C, a 25-km spool of SMF-28 fiber is added to the
7-km link. With the additional loss of the spool (factor of 500 at
1064-nm each way) the reflection method is no longer practical, and
only the transceiver method is characterized for the 32-km link. No
special care is taken to isolate the spool from its environment. For
all three setups, the first and last 5-m sections of fiber are
single-mode for 1064-nm light in an effort to mitigate any effects
of time-changing transverse mode dispersion from the slightly
multi-mode BRAN fiber.  Attenuation through 7~km of the fiber at
1064~nm is measured to be $< 4$~dB. All but one interconnection
along the 7-km fiber path is fusion spliced to reduce unwanted
reflections, leading to the improvement over previous
results~\cite{ye03}.

The out-of-loop measurements are made using several different
heterodyne beats. We adopt a notation in writing the beat
frequencies as
$f_{\mathrm{loc,rem}}^{\left(\mathrm{dist.}\right)}$, where the
first and second subscripts describe light at the local and remote
ends, respectively.  The superscript denotes the length (km) of
the stabilized fiber link. Explicitly, the beat between the local
transfer laser and the remote light exiting the gold-coated fiber
is written as $f_{\mathrm{tfr,tfr}}^{\left(7\right)}$, and
$f_{\mathrm{tfr,rpt}}^{\left(7\right)}$ denotes the beat between
the local transfer laser and the repeater light from the remote
transceiver laser. Similarly, the beat between a mode of the local
frequency comb and the remote light exiting the gold-coated fiber
is $f_{\mathrm{fs,tfr}}^{\left(7\right)}$ while
$f_{\mathrm{fs,rpt}}^{\left(7\right)}$ denotes the local comb's
beat frequency against the remote repeater's light for the 7-km
link. For the 32-km link, only
$f_{\mathrm{fs,rpt}}^{\left(32\right)}$ is characterized.
Measurements against the local transfer laser give information
only about the fiber link, whereas measurements against the local
frequency comb include the phase lock between the transfer laser
and frequency comb.

\begin{figure}[t]
\resizebox{8.5cm}{!}{
\includegraphics[angle=0]{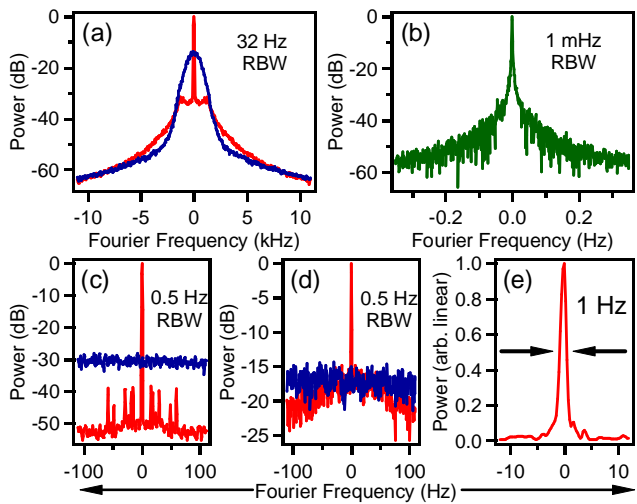}}
\caption{\label{Fig2}(color online) Power spectra of various
out-of-loop heterodyne beats. (a)
$f_{\mathrm{tfr,tfr}}^{\left(7\right)}$ characterizes the 7-km
transfer only (b) $f_{\mathrm{fs,tfr}}^{\left(7\right)}$
additionally characterizes the transfer laser's lock to the local
fs comb. (c) same as (a), at finer resolution. (d)
$f_{\mathrm{fs,rpt}}^{\left(32\right)}$ characterizes the 32-km
transfer as well as the transfer laser's lock to the local fs
comb.  (e) $f_{698,1126}$ is the effective beat between
independent clock lasers separated by 3.5 km and 163 THz. For (a),
(c), and (d) red is for the actively-stabilized network and blue
is for the passive case.}
\end{figure}
Figure~\ref{Fig2} summarizes the results of linewidth
characterization for the various measurements.
Figure~\ref{Fig2}(a) displays the power spectrum of
$f_{\mathrm{tfr,tfr}}^{\left(7\right)}$ to characterize the 7-km
fiber transfer. The uncancelled fiber noise broadens the
transferred linewidth to $\sim$1 kHz, while the coherent narrow
peak is achieved under active fiber noise cancellation, with a
1.3-kHz servo bandwidth limited by $T_{rt}$ for the 14-km round
trip. Figure~\ref{Fig2}(c) shows the same data at finer
resolution; the energy that spread into a 1-kHz bandwidth by the
passive fiber's phase noise is replaced into the narrow central
carrier under active noise cancellation. Figure~\ref{Fig2}(b)
shows $f_{\mathrm{fs,tfr}}^{\left(7\right)}$ as a way to
additionally characterize the transfer laser's lock to the local
frequency comb; the full system operates with a sufficiently small
phase noise that a 1-mHz linewidth is recovered.
Figure~\ref{Fig2}(d) shows $f_{\mathrm{fs,rpt}}^{\left(32\right)}$
with active noise cancellation of the 32-km link (narrow carrier
in red) and without noise cancellation (flat noise in blue). A
narrow coherent feature is present at 0.5-Hz resolution bandwidth,
but with less signal-to-noise than for the 7-km transfer because
the servo bandwidth is smaller for the longer link. Finally,
Fig.~\ref{Fig2}(e) shows the effective heterodyne beat between the
two stable clock lasers at JILA and NIST, linked by the 3.5-km
noise-cancelled fiber and two independent optical combs. The whole
system preserves the full phase coherence of our optical frequency
references.

\begin{figure}[t]
\resizebox{8.0cm}{!}{
\includegraphics[angle=0]{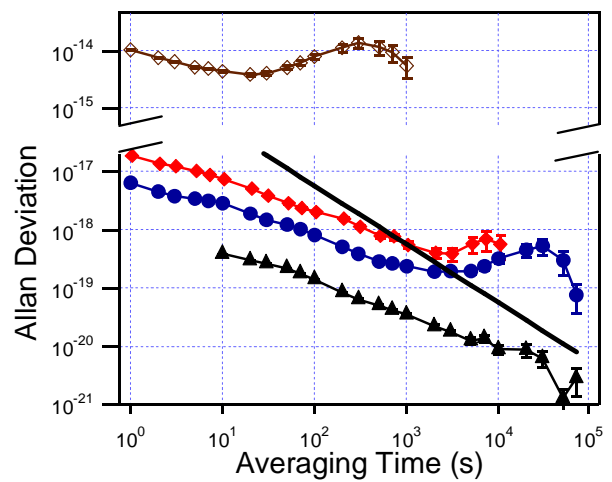}}
\caption{\label{Fig3}(color online) Instability of the transfer
systems. Open brown diamonds are typical passive instabilities of
the 7- and 32-km fiber links. Closed red diamonds are for the 32-km
system, measured by frequency counting
$f_{\mathrm{fs,rpt}}^{\left(32\right)}$. Closed blue circles are for
the 7-km fiber transfer, using
$f_{\mathrm{tfr,tfr}}^{\left(7\right)}$.  Black triangles are the
instability of $f_{\mathrm{fnc}}$ used for in-loop noise
cancellation. The solid black line represents 1 radian of
accumulated phase noise during the averaging time.}
\end{figure}
We also directly count the frequencies of the various out-of-loop
heterodyne beats. For improved counting resolution, we mix the
beat frequencies to 10~kHz using a sufficiently stable rf source.
Figure~\ref{Fig3} displays the resultant Allan deviations. The
open diamonds (in brown) show a typical passive instability for
both the 7- and 32-km links. Closed diamonds (in red) are from
counting $f_{\mathrm{fs,rpt}}^{\left(32\right)}$ to measure the
instability of the 32-km transceiver system. Closed circles (in
blue) use $f_{\mathrm{tfr,tfr}}^{\left(7\right)}$ to measure the
instability of the 7-km fiber transfer (reflection method). Closed
triangles (in black) are the instability of $f_{\mathrm{fnc}}$
used for active stabilization of both the 7- and 32-km systems,
divided by 2 for a fair comparison against the one-way out-of-loop
instabilities. Also shown is a solid black line indicating the
effect of 1 radian of accumulated phase noise during the averaging
time; the 7-km measurement lies below this level for averaging
times $> 10^3$ seconds. The 7-km fiber transfer (reflection
method) is locked continuously for 70 hours, and for frequency
counting $f_{\mathrm{tfr,tfr}}^{\left(7\right)}$ with a 10-s gate
time, only $0.3\%$ of the counts are outliers. The transfer's
accuracy (residual offset from the expected frequency) is $1
\times 10^{-19}$, consistent with the long-timescale instability.
The 32-km system is continuously locked for several hours at a
time for 14 hours (net $85\%$ duty cycle), with a similar amount
of outliers.  Brief interruptions are caused by the clock laser or
the fs frequency comb losing lock, not the transfer or transceiver
lasers or the fiber noise cancellation.  The accuracy is $5.2
\times 10^{-19}$, again consistent with the measured instability
from the diurnal temperature fluctuation.  To achieve $10^{-17}$
($10^{-19}$) accuracy at 1-s ($10^5$-s), the various $\sim100$-MHz
rf frequency references used for frequency offsets and phase locks
must be accurate to $\sim3 \times 10^{-11}$ ($3 \times 10^{-13}$).

Both the 7- and 32-km measurements are limited by fluctuations of
the out-of-loop optical components used to mix the various optical
frequencies.  This is confirmed by replacing the 7-km fiber with a
2-m fiber, yielding identical results for timescales up to $10^4$
s, even though the passive instability of the 2-m fiber is 100
times smaller than for the 7-km fiber. The system is enclosed in a
box to isolate the out-of-loop optics from air currents and
acoustic pickup; without the box the 1-s instability rises to
several parts in $10^{16}$.  The excess instability at $2 \times
10^4$~s is caused by daily temperature drift of the laboratory.
The 32-km measurement is a factor of 2.5 worse than the 7-km
measurement at all averaging times, suggesting that the limitation
is still the out-of-loop measurement system with greater
instability due to its larger complexity. This fact is consistent
with a comparison of the instabilities of the 7-km and 32-km
transceiver links (measured by counting
$f_{\mathrm{tfr,rpt}}^{\left(7\right)}$ and
$f_{\mathrm{tfr,rpt}}^{\left(32\right)}$, respectively) which are
identical for averaging times up to 1000 s.

\begin{figure}[t]
\resizebox{\linewidth}{!}{
\includegraphics[angle=0]{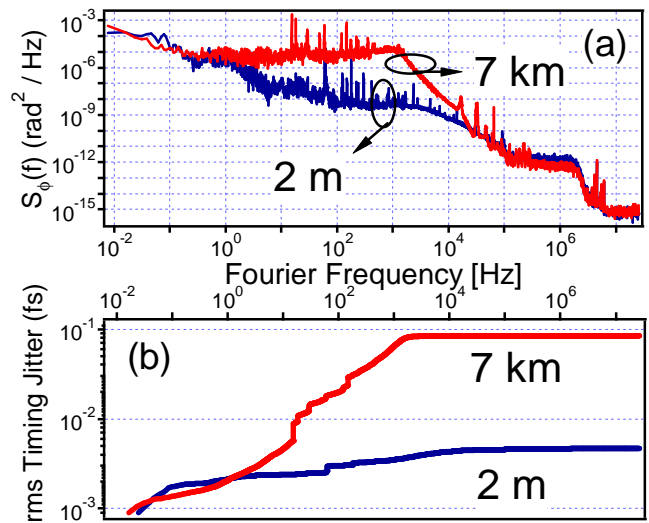}}
\caption{\label{Fig4}(color online) (a) Relative phase noise
spectral density between the local cw transfer laser and light
exiting the 7-km fiber link at the remote end.  (b) The relative
phase noise shown in (a) is integrated and displayed as rms timing
jitter, demonstrating 7-km timing transfer with 0.08 fs of jitter
integrated from 8 mHz to 25 MHz.}
\end{figure}
The fiber transfer's coherence is revealed by a direct measurement
of the phase noise it introduces. Figure~\ref{Fig4}(a) shows the
phase-noise spectral density $S_{\phi}\left(f\right)$ of the 7-km
fiber network (upper curve in red), measured by comparing the
phase of $f_{\mathrm{tfr,tfr}}^{\left(7\right)}$ against a
phase-stable rf reference.  The lower curve (in blue) shows the
same measurement when the 7-km fiber is replaced by the 2-m fiber.
For Fourier frequencies between a few Hz and the 1.3-kHz servo
bandwidth, the active noise cancellation of the 7-km fiber does
not achieve the measurement noise floor represented by the 2-m
data. However, below a few Hz the out-of-loop measurement scheme
dominates the noise, and is common to both lengths of fiber.  The
integrated phase noise is displayed as rms timing jitter in
Fig.~\ref{Fig4}(b). Integrated from 8 mHz to 25 MHz, only 0.085 fs
of timing jitter is accumulated, corresponding to 0.15 rad at the
optical transfer frequency of 282 THz.

We acknowledge funding support from ONR, NIST, and NSF. We are
grateful to J. Bergquist, T. Schibli, D. Hudson, S. Blatt, M. Boyd,
T. Zelevinsky, and G. Campbell for technical help and discussions,
and N. Newbury for loan of equipment.

\end{document}